\documentclass[twocolumn,10pt]{article}
\usepackage[a4paper,margin=1.9cm]{geometry}
\usepackage[utf8]{inputenc}
\usepackage[sort&compress,square,numbers]{natbib}
\usepackage{graphicx}
\usepackage{amsmath,amssymb}
\usepackage{braket}
\usepackage{xcolor}
\usepackage{url}
\usepackage{multirow}
\usepackage{hyperref}
\usepackage{cleveref}
\usepackage[utf8]{inputenc}
\DeclareUnicodeCharacter{2212}{-}   % Minus sign
\DeclareUnicodeCharacter{2013}{--}  % En dash
\DeclareUnicodeCharacter{2014}{---} % Em dash
\DeclareUnicodeCharacter{00A0}{ }   % Non-breaking space

\title{Cathodoluminescence Enhancement Mechanisms in Silica Microspheres}

\author{
Hadar Aharon$^{1}$,
Zahava Barkay$^{2}$,
Sophie Meuret$^{3}$,
and Ofer Kfir$^{1}$\\[5pt]
\small $^{1}$School of Electrical and Computer Engineering, Iby and Aladar Fleischman Faculty of Engineering,\\ \small Tel Aviv University 69978, Tel-Aviv, Israel\\
\small $^{2}$Jan Koum Center for Nanotechnology and Nanoscience, Tel-Aviv University 69978, Tel Aviv, Israel\\
\small $^{3}$CEMES–CNRS, 31055 Toulouse Cedex 4, France
}

\date{}

\begin{document}
\twocolumn[
\maketitle
\begin{abstract}
Cathodoluminescence (CL) enables optical-frequency analysis of samples with nanometer resolutions, originating from the interaction of a focused electron beam with radiative electronic states, or directly with the optical modes of the sample. Here we decompose the various mechanisms underlying CL generation and emission from an archetype spherical resonator using its spectral, angular and spatially resolved features. We investigate radiation of optical whispering-gallery modes in regimes of coherent and incoherent luminescence. The use of different experimental regimes allows us to disentangle the different contributions to the CL in spheres, namely, photon absorption, generation and radiative leakage, and conclude that the photon generation occurs precisely on the sphere's surface. In addition, the spheres serve as high-NA collimating lenses for CL, resulting in mode quality unprecedented for CL in free space. We believe that such collimated and directed CL in free space will enhance existing quantum measurements of CL and facilitate new ones, such as high-rate electron-photon entangled pairs, CL from quantum emitters, and homodyne analysis of CL.
\end{abstract}

\vspace{2em}
]

\maketitle

\section{Introduction}
Electron microscopes are ubiquitous analytical tools for the ultrasmall, enabling imaging, diffraction and spectroscopy of materials down to the atomic scale \cite{shindo2013,pennycook2011,goodhew2000}. The ability to pinpoint the electron beam with an accuracy well below an optical wavelength is used to quantify local optical responses that would otherwise be smeared by the light diffraction limit \cite{polman2019,MEURET2019,KOCIAK2017,Bendana2011}, such as the mapping of evanescent fields, dark modes \cite{Barrow2014,srinivasan2005}, high momentum states \cite{li2021,krehl2018,shekhar2017}, and more. Capturing photons emitted from an impinging electron, namely cathodoluminescence (CL), was applied to characterize plasmonic nanocavities and surface plasmon polaritons \cite{elibol2025plasmonic,zar2022,Takumi2020,Ron2020,Knight2015,garcia2013,Kuttge2009}, resonant dielectric nanostructures \cite{Soler2025,Auad2021,Mignuzzi2018}, mineral coloration in geoscience \cite{gotze2002,pagel2000}, thin films in semiconductors \cite{romero2003}, etc.; providing access to their spectral, spatial, polarization, and angular-resolved emission properties \cite{Ebel2025,Coenen2011,akerboom2025angle}. An alternative approach for probing photon generation instead of emission is EELS \cite{bezard2024, Losquin2015}, which stands for electron energy loss spectroscopy, used in a complementary manner to probe the correlations and entanglements formed by the electron-photon coupling \cite{varkentina2022, Feist2022,preimesberger2025,henke2025,kazakevich2024,tziperman2025two}.  

The high resolution of CL is particularly powerful for analyzing the properties of discrete modes within optical resonators. It probes the local density of states of photonic crystals and defects, both spatially and angularly \cite{bezard2024,lingstadt2023,Mignuzzi2018,Osorio2016,sapienza2012}, one-dimensional pillar resonators \cite{mignuzzi2024}, and optical whispering-gallery modes (WGMs) circulating on the edge of a dielectric \cite{HeebnerJohnE2008,Auad2021,Muller2021,Feist2022,Machfuudzoh2023,adachi2025temperature}. Such modes are typically characterized by global properties as their quality factor (Q-factor, or Q), modal volume (V), polarization (TM or TE, for transverse magnetic or electric polarization), radiation loss and absorption \cite{Joannopoulos2008, HeebnerJohnE2008}. But in the context of their interaction with an electron and the generation of CL, local properties become important, as well as the particular coupling mechanism. One can divide the electron-sample interaction into coherent and incoherent processes. Coherent CL stems from integrating over the electron trajectory in space and time in a manner that matches the dynamics of the optical mode's field, sometimes referred to as phase matching. It includes transition radiation \cite{Ebel2025,MENDIS2016,Coenen2011,Talebi2019}, Cherenkov radiation \cite{arend2024,adiv2023,gong2023,sapienza2012,taleb2023phase}, and Smith-Purcell radiation \cite{massuda2018,remez2017}. In incoherent CL the generated charge carriers populate intrinsic material excitations, such as defects, excitons, and higher bands \cite{Xu1995,ye2008,wade2004}, etc., followed by spontaneous emission \cite{Mignuzzi2018}, for which the phase is undetermined. In practice, the generated CL photons need to reach a detector, which typically requires their out-coupling in its direction, which complicates the analysis of optical cavities. On one hand, a good resonator may have a high Q-factor and density of states, but the confinement of light to it inherently means that radiation to the environment is inhibited. On the other hand, a poor resonator would emit broadly spectrally and angularly, suffering from a low signal-to-noise ratio, and may be dominated by other optical signals of the sample. These are particularly harsh tradeoffs for quantum optics, where one ideally has a narrowband emission with a well-defined radiative mode \cite{ruimy2025,kfir2021,lvovsky2009,hansen2001,grzesik2025quantum}.

In this work we investigate the CL properties and enhancement mechanisms of individual silica microspheres. We use a focused electron beam in a scanning and in a transmission electron microscope (SEM and TEM) with beam energy of 30 keV and 200 keV, respectively, and use spheres with several diameters, 2.1 $\mu m$, 4.4 $\mu m$ and 62 $\mu m$. We show that different photon generation and radiation phenomena are separable spatially and angularly. Varying the electron beam energy and the sphere's diameter disentangles CL generation mechanisms, radiative efficiency, and material absorption. As an example for competing luminescence phenomena, we address the collimation of external CL, showing that its dominance under some parameters may hinder the analysis or be used as a beneficial feature for out-coupling CL with a good spatial mode.

\begin{figure}[!htbp]
\centering
\noindent
\includegraphics[width=\columnwidth]{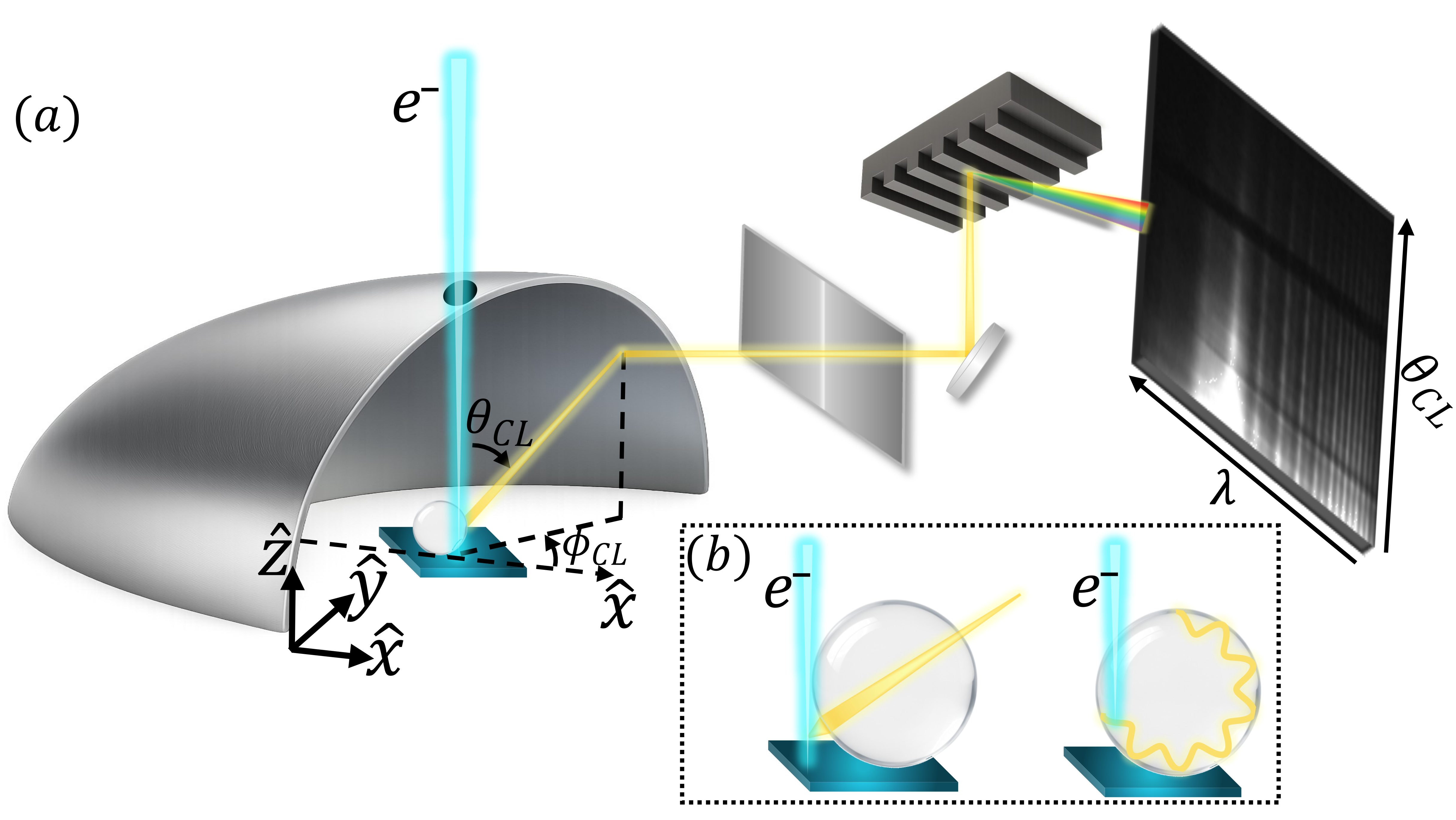}
\caption{Schematic of the measurement setup. (a) An electron beam is focused onto the sample, generating CL photons, which are collimated through a parabolic mirror. A slit before the grating maps the spectrum vs. the polar angle $\theta$, around $\phi=0^\circ$. (b) The inset shows CL enhancement by excitation of a resonant WGM (right) and by geometrical beam collimation (left).} 
\label{fig:illustration}
\end{figure}

\section{Results}
We use an Apreo 2S-LV SEM by Thermo Fisher Inc., retrofitted with the SPARC CL analyzer and a parabolic collection mirror by Delmic Inc. as our main experimental platform (see Fig.\ref{fig:illustration}). The parabolic mirror projects the angular distribution of the CL directly onto the camera. Alternatively, a $\theta$-resolved spectrum is captured by placing a vertical slit parallel to the azimuthal angle $\phi=0^\circ$. The polar angle $\theta$ is mapped on the vertical axis of the camera (positive values for $\phi=0^\circ$ and negative values for $\phi=180^\circ$), and the wavelength spans the horizontal axis.

\begin{figure}[!htbp]
\centering
\noindent
\includegraphics[width=\columnwidth]{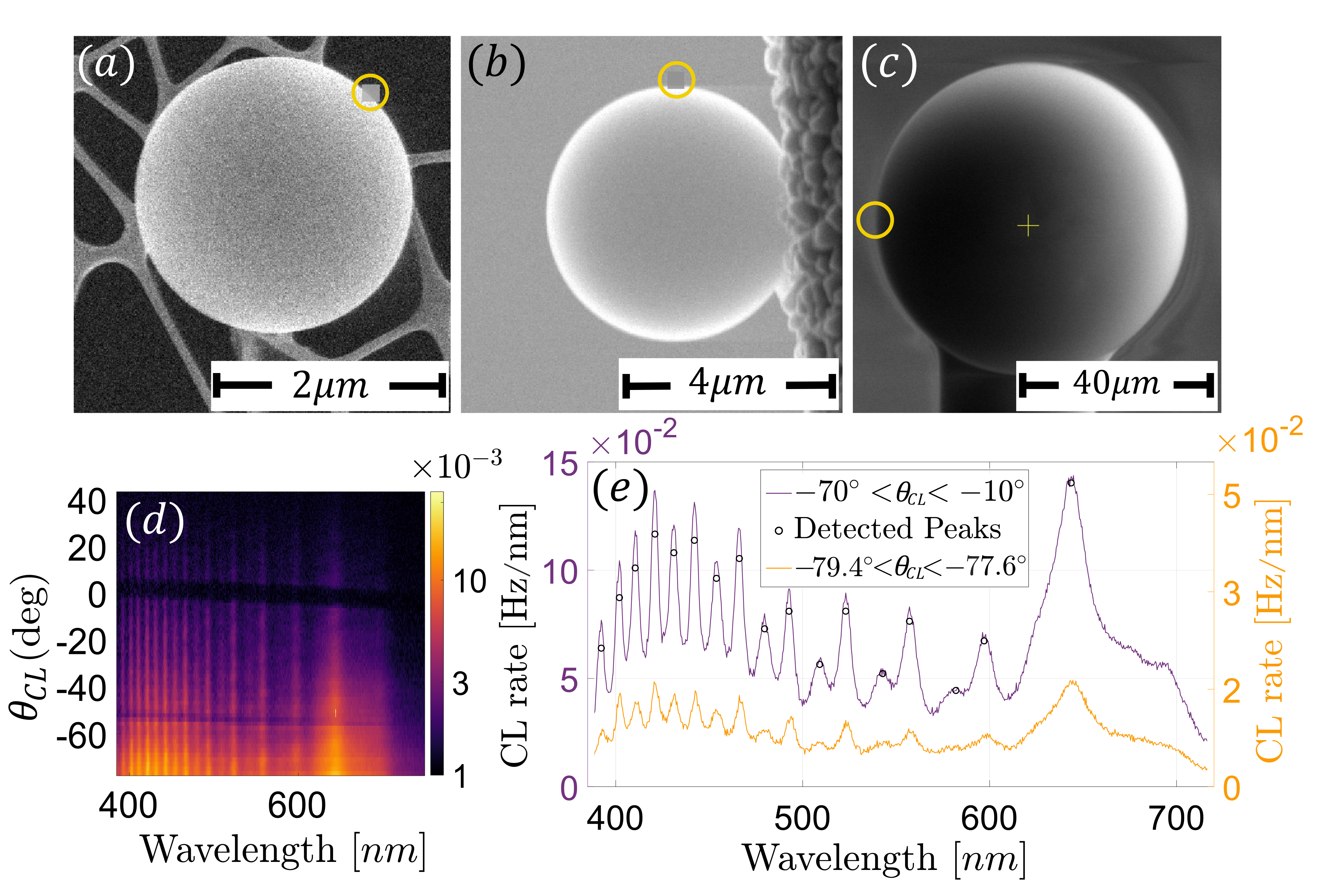}
\caption{Angular-resolved CL spectrum. (a)–(b) SEM images of silica microspheres with diameters of 2.1 $\mu m$ , 4.4 $\mu m$, respectively. The electron-beam positioning is marked by a gray square circled in orange. (c) SEM image of a 62 $\mu m$ diameter sphere fabricated from a tapered optical fiber. (d) $\theta$-resolved CL spectrogram from the 2.1 $\mu$m sphere, showing peaks of the  sphere's WGMs with a spatially varying background. (e) CL spectral count rate extracted from (d), summed over a low-background range  $-70^\circ\leq\theta_{CL}\leq-10^\circ$ (purple) and a high-background range $-79.4^\circ \leq \theta_{CL} \leq -77.6^\circ$ (orange)}
\label{fig:AR-spectrum}
\end{figure}

We analyze CL generation and emission mechanisms in silica glass microspheres with diameters of 2.1 $\mu$m, 4.4 $\mu$m, and 62 $\mu$m, as shown in the secondary-electron micrographs in Fig.\ref{fig:AR-spectrum}(a)–(c), acquired in the SEM at an acceleration voltage of 30 keV. The electron beam positioning is marked by a gray square circled in orange in Fig.\ref{fig:AR-spectrum}(a)-(c). The 62 $\mu$m sphere, fabricated from a tapered optical fiber, exhibited no clear resonances in the CL and spectra from the 4.4 $\mu m$ sphere are discussed later. The smaller two are silica microspheres (Bangs Labs, SSD5003 and SSD4002), provided as a dry powder. The powder was dispersed in isopropanol and drop-cast onto a standard 3 mm TEM grid coated with a 20 nm thick carbon film and laces, respectively (TED Pella 01841 and 01894). Fig.\ref{fig:AR-spectrum}(d) presents a $\theta$-resolved spectrum acquired with an integration time of 30 seconds from the 2.1 $\mu$m sphere. The dark horizontal stripe at $\theta_{CL}=0$ arises from a hole in the mirror through which the electron beam passes. The purple spectrum in Fig.\ref{fig:AR-spectrum}(e) is integrated over the angular range $-70^\circ \leq \theta_{CL} \leq -10^\circ$ , which is the widest range within the low background region of the spectrogram. The orange spectrum is from the range  $-79.4^\circ \leq \theta_{CL} \leq -77.6^\circ$, in which competing luminescence phenomena decrease the peak contrast. Across the acquisition band, we identify 17 distinct spectral peaks corresponding to WGMs, nearly independent of the emission angle. The WGM signal dominates in the darker regions of the angular distribution. Therefore, the absence of strong angular features is called "WGM-regime". 

\begin{figure}[!htbp]
\centering
\noindent
\includegraphics[width=\columnwidth]{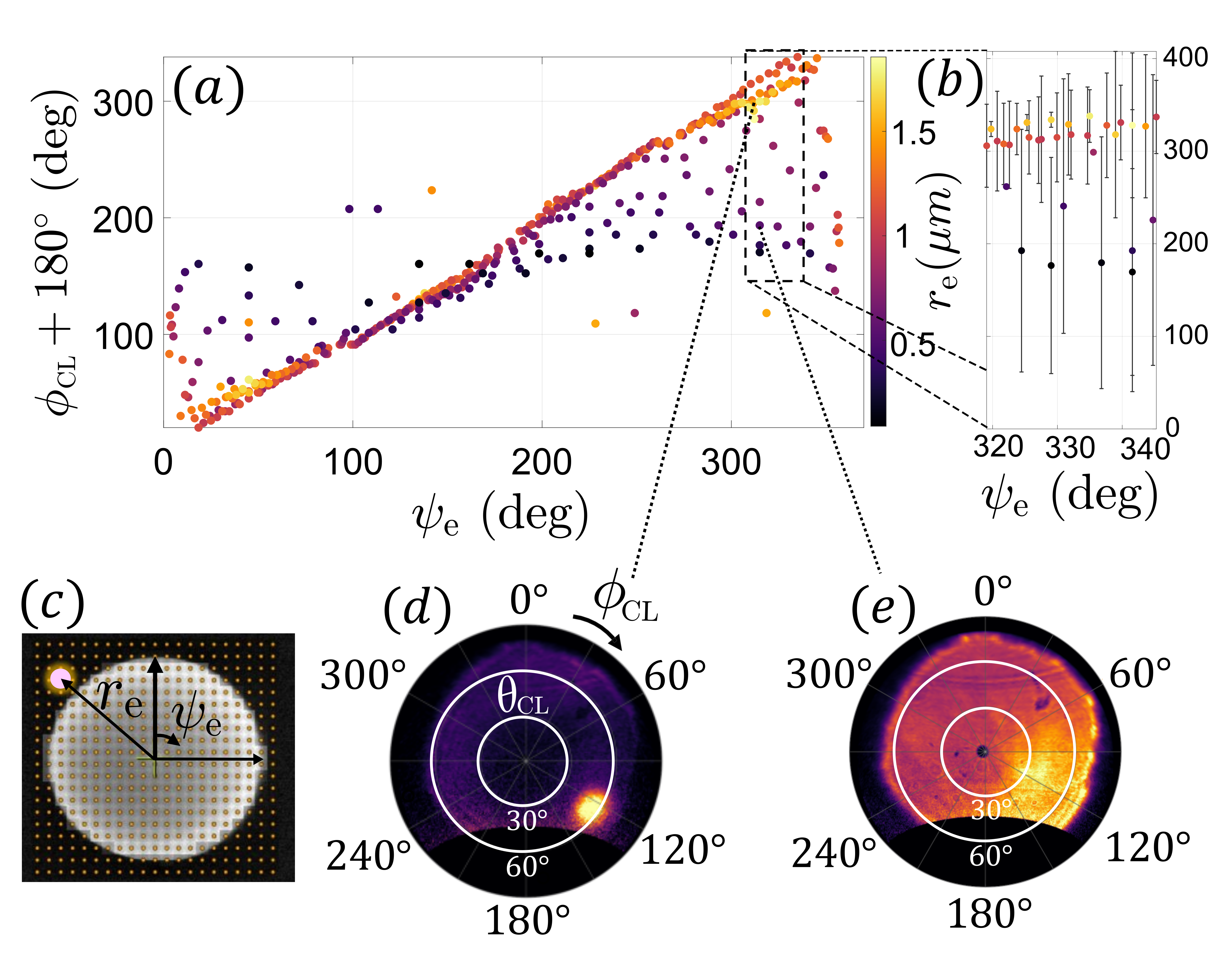}
\caption{Angular correlations of CL emission with electron impact coordinate. (a) CL emission peak angle $\phi_{CL}$ as function of the beam impact angle $\psi_e$, shows an emission trend to the opposing angle due to the rotation invariance and the sphere.  (b) Expanded view of the marked segment, with added error bars. (c) A grid of impact positions of the electron beam, overlaid on a 2.1 $\mu m$ sphere.  (d)-(e) 2D maps showing the CL angular distribution for electron excitation at $\psi_e = 320^\circ , r_e = R_{sphere}$ (d) and $\psi_e = 320^\circ, r_e = 0.2 R_{sphere}$ (e).}
\label{fig:phi_CL}
\end{figure}

\begin{figure}[!htbp]
\centering
\noindent
\includegraphics[width=\columnwidth]{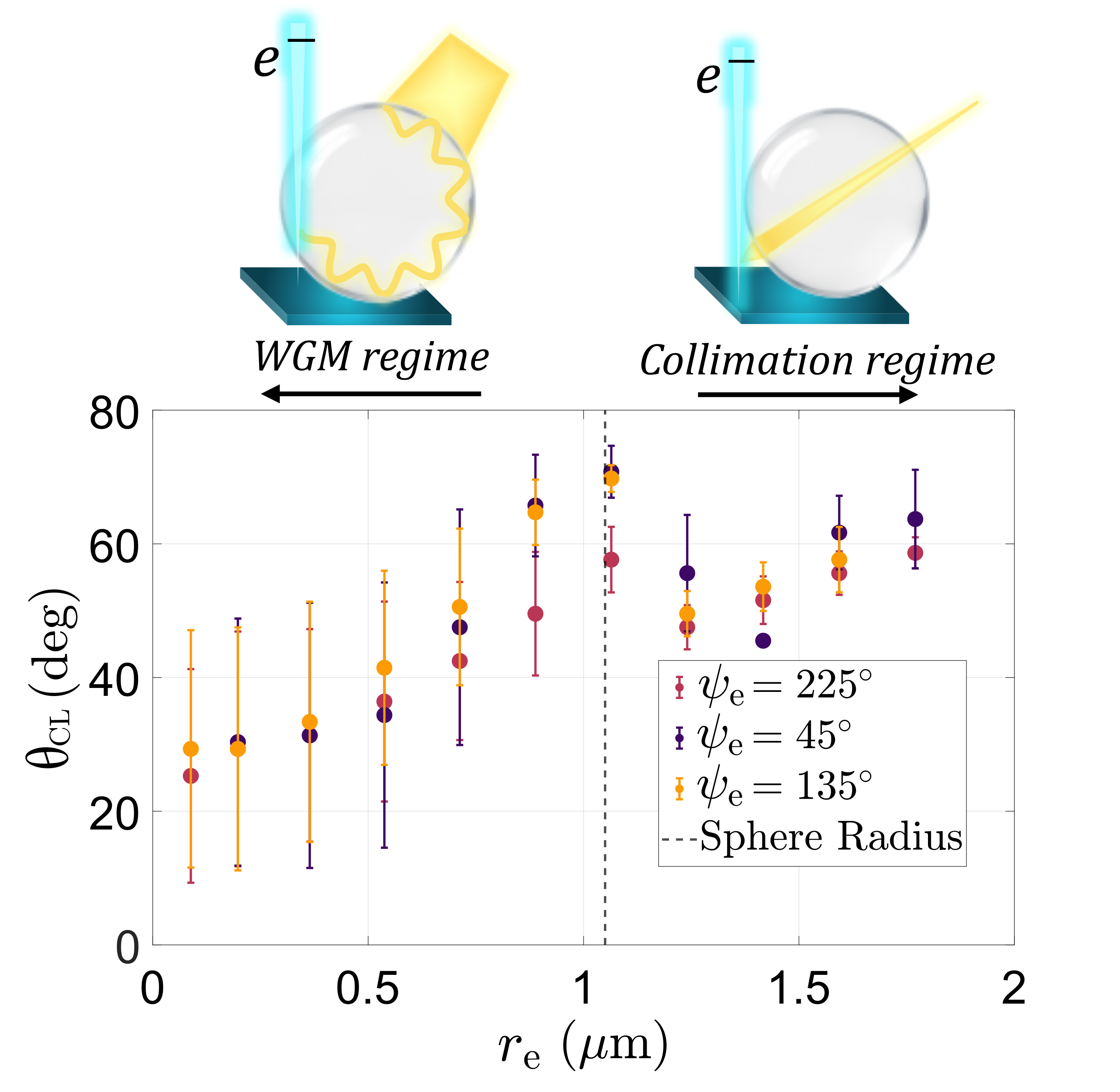}
\caption{Detected CL emission peak angle $\theta_\text{CL}$ as function of the electron beam radial distance from the center of the sphere $r_e$, shown for three different $\psi_e$ angles. The sphere radius is indicated by the black dashed line. }
\label{fig:theta_CL}
\end{figure}

Figure \ref{fig:phi_CL} shows the role of the electron-beam's impact parameter on the CL distribution from the 2.1 $\mu$m microsphere, integrated over the entire spectrum. The spatial properties of the CL are analyzed in four-dimensions, two for the electron incidence and two for the emitted CL. Each position in the impact plane (see Fig.\ref{fig:phi_CL}(c)) is defined by the impact parameter, $r_e$, and angle, $\psi_e$, such that $\vec{r}=r_e(cos(\psi_e)\hat{x} + sin(\psi_e)\hat{y})$. The angular distribution of the CL is mapped by the polar and azimuthal angles, $\theta_{CL},\phi_{CL}$ (see Fig.\ref{fig:illustration}). Fig.\ref{fig:phi_CL}(a) presents $\phi_{CL}$ vs. the impact angle, $\psi_e$. We consider $\psi_e$ as an accurate parameter, whereas for $\phi_{CL}$, we present its mean. Since the physical system is rotationally invariant the emission angle depends linearly on the impact angle, $\phi_{CL}\propto\psi_e$. Here, the dominant feature is that they are diametrically opposed, $\psi_e - \phi_{CL}=180^\circ$, due to the positive curvature of the sphere. A concave surface would overlap the impact and emission angles. For small impact radius $r_e$ (dark data points), the emission angles are poorly defined. It is visually evident in the spread of darker vs. lighter data points, as well as in the standard deviation of $\phi_{CL}$, presented as error bars for a segment of the data in Fig.\ref{fig:phi_CL}(b).
Fig.\ref{fig:phi_CL}(d-e) show the angular CL maps for two exemplary impact positions. Fig.\ref{fig:phi_CL}(d) is the emission from an electron impact on the sphere's perimeter ($r_e\approx R_{sphere}$) and resides on the linear trend $\psi_e-\phi_{CL}=180^\circ$ in Fig.\ref{fig:phi_CL}(a). The CL is nearly collimated, narrowly distributed in both $\phi_{CL}$ and $\theta_{CL}$. In contrast, Fig.\ref{fig:phi_CL}(e) shows the CL angular map for a small impact parameter ({$r_e=0.2R_{sphere}$}), yielding broad and poorly defined emission angles. The corresponding data points in Fig.\ref{fig:phi_CL}(a) deviate from the diametrically opposed trend. The angular-independent background CL is more visible in Fig.\ref{fig:phi_CL}(e) than in \ref{fig:phi_CL}(d), since they are normalized to the highest value. That is, Fig.\ref{fig:phi_CL}(e) represents the WGM-regime, where the angularly localized CL phenomenon is not dramatically stronger. The polar angle $\theta_{CL}$ is analyzed in Fig.\ref{fig:theta_CL}. There is a monotonic trend for $\theta_{CL}$ vs. $r_e$ up to the sphere's edge. In principle, one would expect that for $r_e\to 0$, $\theta_{CL}$ would nullify, but the CL distribution is too wide to show that clearly. However, for $r_e>R_{sphere}$, the angular distribution shrinks, forming a sharp spatial mode as in Fig.\ref{fig:phi_CL}(d), and the monotonicity with $r_e$ stops. Therefore, the two regimes can be clearly distinguished: $r_e < R_{sphere}$ and $r_e > R_{sphere}$ as the WGM-regime and collimation-regime, respectively.

\section{Discussion}
The competition between CL processes can be addressed by utilizing the distinct degrees of freedom that distinguish them and analyzing each process in the regime where it dominates. Approaching in a naive manner, capturing a spectrum without spatial filtration would mix directed CL and omnidirectional CL, such as from WGM, whereas spatial filtration as in Fig.\ref{fig:AR-spectrum}(d) may isolate the weaker effect. When the electron hits the edges of the sphere, spatial filtration becomes key. Without spatial selectivity the signal is dominated by the collimated output. Suppressing by a slit or an aperture would elevate the signal of competing effects. In fact, the on-sample collimation enabled by the silica sphere could be a beneficial feature for downstream applications of the CL. First, it collects a solid angle of 1.84 $sr$, centered at an inclination around $45^\circ$ (see illustration in Fig.\ref{fig:illustration} and data in Fig.\ref{fig:phi_CL}(d)). For an omnidirectional emission the collection efficiency would then be $1.84/{4\pi}=14.6\%$. Second, the outgoing mode has a low-divergence, with nearly a transform-limited Gaussian profile. The measured standard deviation of the angular distribution in Fig.\ref{fig:phi_CL}(d) is $\sigma_\theta\approx \sigma_\phi=5^\circ$. Assuming that the mode has an initial size equal to the sphere's diameter, we estimate that its emittance is 0.17 $\mu m \cdot rad$, corresponding to $M^2 \approx 2$ \cite{yariv2007}. For some context, consider the constraints of the CL-collection mirrors in TEMs, which protrude into the narrow objective-lens gap  \cite{Zagonel2012,Tizei2013,Meuret2020,Meuret2021}. The collimation enables the suppression of possible aberrations in the mirror, resulting in a mode with sufficient purity for quantum measurements, or an efficient coupling to a single-mode fiber \cite{kfir2021,chonan2014}.

The opposing regime, in which the CL spreads generally evenly in space, is the WGM regime in the case of the sphere microresonators, and we turn to an angular-independent analysis to disentangle its features. The CL signal inherently convolves the contributions of the (i) generation of a photon in the resonator, (ii) radiative coupling to free space, and (iii) material absorption. To identify these components separately, we use spheres with several diameters, and tune the electron-beam energy. We begin by addressing the competition between the loss mechanisms, namely, photon absorption and radiation to free space. That is done through their manifestation within the quality factor of the microresonator. After addressing the loss, we analyze the generation processes of light in the resonator. 

Let us consider the processes following the population of a WGM, regardless of how it was excited. The interplay between the material absorption and radiation is reflected in their respective lifetimes, $\tau_{mat}$ and $\tau_{rad}$. These are typically expressed in terms of the quality factors,  $Q_{mat}=\omega\tau_{mat}$ and $Q_{rad}=\omega \tau_{rad}$, where $\omega$  is the optical angular frequency of a chosen spectral peak. The total quality factor, which is measurable, is then
\begin{equation}
\frac{1}{Q_{tot}}=\frac{1}{Q_{rad}}+\frac{1}{Q_{mat}} ,
\label{eq:eq1}
\end{equation}
The balance between the decay mechanisms varies with the sphere's size. As the diameter increases from 2.1 $\mu m$, through 4.4 $\mu m$ to 62 $\mu m$ (Fig.\ref{fig:AR-spectrum}(a)-(c)) the radiative loss diminishes. Thus, the balance between the radiative and absorptive mechanisms in the 2.1 $\mu m$ spheres shifts to an absorption-dominated regime in the 4.4 $\mu m$ spheres, whereas the CL in the largest spheres is entirely suppressed. To give a quantitative example, the theoretical radiative lifetime for TM-polarized modes is 28 fs and $10^4$ fs in the 2.1 and 4.4 $\mu m$ microresonators, respectively. These values are calculated for a similar wavelengths, 503 $nm$ and 505 $nm$. The lifetime is the inverse linewidth calculated using the WGMode package for Matlab \cite{BALAC2019}, for a given refractive index (1.45) and the diameters mentioned above. For all the identifiable peaks in the spectrum, the quality factor of a peak centered at a wavelength $\lambda$ is extracted from its full-width-at-half-maximum (FWHM) bandwidth, $ \Delta \lambda$, by $Q_{tot}=\lambda \Delta \lambda^{-1} $ \cite{HeebnerJohnE2008}.

\begin{figure}[h]
\centering
\noindent
\includegraphics[width=\columnwidth]{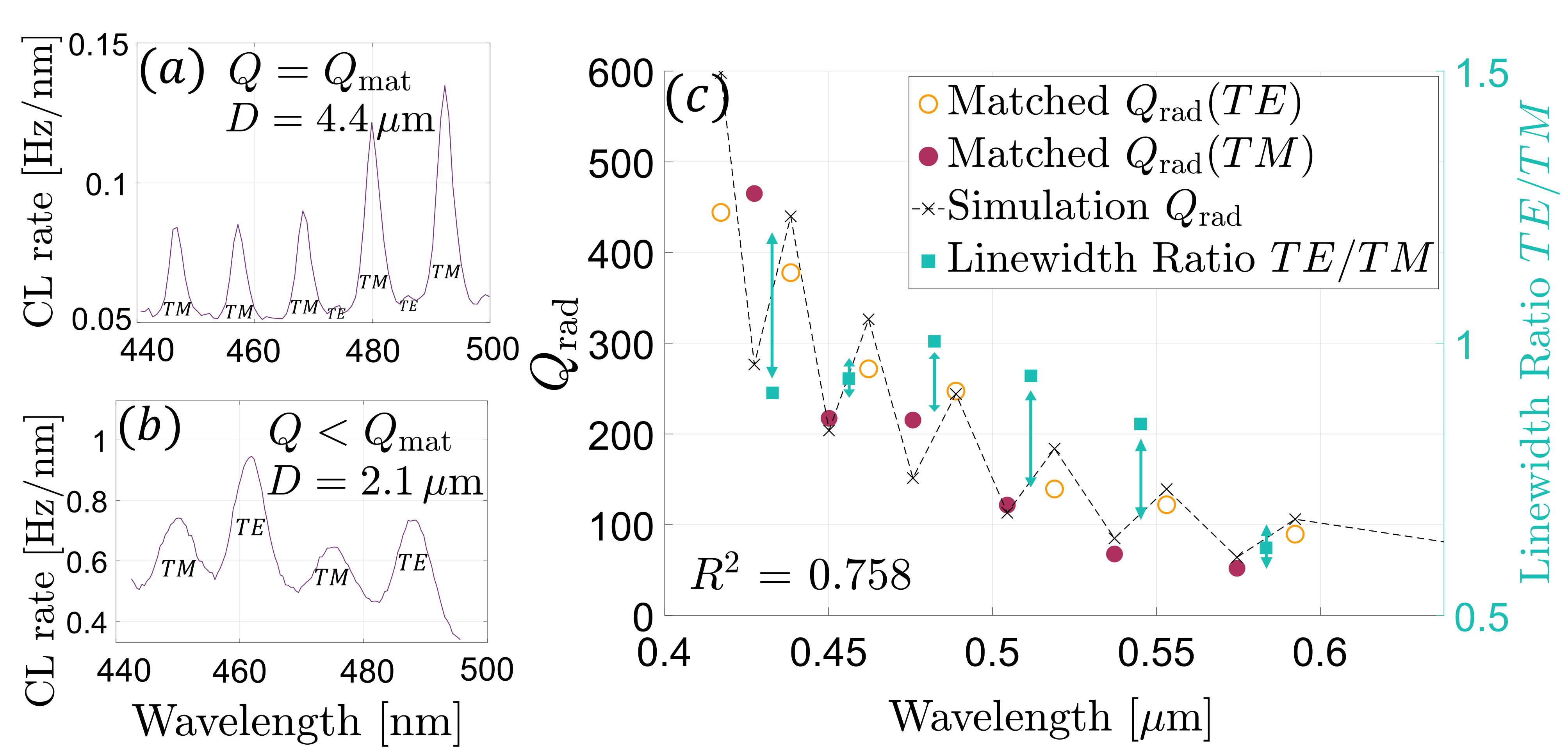}
\caption{Quality factor decomposition of silica microspheres. (a) Spectrum of the 4.4 $\mu m$ sphere and (b) 2.1 $\mu m$ sphere. The linewidth of the peaks in (a) was used to extract the material-loss quality factor, $Q_{mat}$. (c) Radiation quality factor, $Q_{rad}$, derived from the experimental linewidth of the peaks in (b) for TE (open circles) and TM polarized (filled circles). The x-markers are the radiative quality factors from the simulated whispering-gallery loss. The dashed line is a guide to the eye. The filled squares (right axis) are the linewidth ratio of TE- vs. TM-polarized modes.}
\label{fig:Qfactor}
\end{figure}

The absorption-dominated regime allows us to focus only on the material contribution to the quality factor. For the 4.4 $\mu m$ sphere, the quality factor extracted from the experimental spectrum in Fig.\ref{fig:Qfactor}(a) is $Q_{tot}\sim 10^2$. Since the calculated radiative component is $Q_{rad} \sim 10^4 $, its contribution to eq.(\ref{eq:eq1}) is indeed negligible, and one can assume that $Q_{tot} \cong Q_{mat}$. Hence, the measured quality factor across the entire spectrum quantifies the absorption. In principle, one can retrieve the radiative component for the 2.1 $\mu m$ sphere (Fig.\ref{fig:Qfactor}(b)) by substituting in eq.\ref{eq:eq1}, the $Q_{mat}$ that was measured in the larger spheres. However, the detected CL in the absorption-dominated regime for an incident electron energy of 30 keV is extremely weak. Furthermore, the spectrum of the 4.4 $\mu m$ sphere is denser than the 2.1 $\mu m$ sphere, resulting in a partial overlap between the TE-polarized and TM-polarized modes (see Supplementary Material, Fig.S1). To overcome this, we use an HF2000 TEM by Hitachi, in which the electrons are accelerated to 200 keV and their path is set to a tangent configuration ($r_e\approx R_{sphere}$) for phase-matching, which increases the generation rate of WGM-photons in the micro-cavity. The CL spectrum from the TEM (see Fig.\ref{fig:Qfactor}(a)) exhibits sharper spectral peaks than in the smaller spheres, with a noticeable propensity for TM-polarization, due to their increased coupling \cite{Auad2021,Kfir2020}  to the traversing electron.  Thus, these well-separated spectral peaks allow an evaluation of $Q_{mat}=Q_{tot}$. For glass microresonators that were prepared in a similar manner, one can interpolate the $Q_{mat}$ based on the analysis of the peaks within the absorption-dominated regime. As a material property, this $Q_{mat}$ should apply across the spectrum and for any polarization. 

In the regime of a balanced radiative and absorptive losses, as for the spheres with 2.1 $\mu m$ diameter, the $Q_{mat}$ measured in the larger spheres can be used within eq.\ref{eq:eq1} alongside the measured $Q_{tot}$ to extract the purely radiative component, $Q_{rad}$. The circle markers in Fig.\ref{fig:Qfactor}(c) show the extracted $Q_{rad}$ for the resonances of the 2.1 $\mu m$ sphere. It exhibits the main expected features. The quality factor decays for longer wavelengths as the angular modes are of a lower order. The TE-polarized modes have a higher quality factor since they are more confined to the sphere, compared with the adjacent TM-polarized modes. The blue squares in Fig.\ref{fig:Qfactor}(c) (right axis) show the ratio between the radiative coupling of the TE-polarized modes vs. TM-polarized modes, which is systematically less than $1$. The black x-markers are the radiative quality factor for the modes of a sphere with 2.1 $\mu m$ diameter with a wavelength-dependent refractive index around $1.45$, simulated using the WGM package. The agreement between the radiative quality factors from our experiment and those from the semi-analytic toolbox is high, with an $R^2 = 0.75$. Above 450 $nm$ the coefficient of determination is far better, reaching $R^2=0.83$.

While typically the CL-photon generation depends on the electron energy, we use the calculated modes to assess that the photons in our experiment form strictly on the surface of the sphere. For the low-energy electrons impinging on silica spheres, the emission is incoherent due to the lack of phase matching. The cathodoluminescence is mediated by material excitations, whose radiation according to Fermi's golden rule is governed by the local density of optical states (LDOS) \cite{Mignuzzi2018,Sola-Garcia2021}, which is the energy density of the WGM, per photon \cite{BALAC2019}. 
The photon generation rate $\Gamma$ is proportional to  
\begin{equation}
\Gamma \propto\frac{|E(\omega,\vec{r})|^2} {|E_{max}|^2}. \frac{\hbar \omega}{n_\lambda^2\epsilon_0  V_\text{eff}} 
\label{eq:photon_generation)rate}.
\end{equation}
$E(\omega,\vec{r})$ is the modal electric field at position $\vec{r}$ and angular frequency $\omega$, $|E_{max}|$ is the maximum value of the Euclidean norm of the electric field $ , n_{\lambda}$ is the refractive index of the mode at wavelength $\lambda$, $\epsilon_0$ is the vacuum permittivity and $V_\text{eff}$ is the effective mode volume.
We compare the generation rate, $\Gamma$, for spectrally adjacent TM-polarized and TE-polarized modes, with central wavelengths of 503 $nm$ and 518 $nm$, respectively. The ratio between the two polarization modes traces-out effects as out-coupling, detector efficiency and spectral response. Thus, it keeps the analysis purely on the photon generation physics and enables us to compare experimental parameters with their prediction from the semi-analytic calculations. 
The measured CL counts from a particular mode, $N_{CL}$, depends on the generation rate and the probability to radiate rather than undergo absorption, $N_{CL} \propto \Gamma\frac{Q_{mat}}{Q_{rad}+Q_{mat}}$. One can see that this form represents the competition between loss and radiation. In the limit of a non-absorptive material, $Q_{mat}\to \infty $, $N_{CL}\to \Gamma $, and all the generated photons radiate. In the opposite limit, $Q_{mat}\to 0 $, $N_{CL}\to 0$. In this limit of a poor material quality, the relative emission between TE and TM depends on their radiative rate, that is, of a photon manages to exit the cavity before it is absorbed, $\frac{N_{CL,TE}}{N_{CL, TM}} = \frac{\Gamma_{TE}}{\Gamma_{TM}}\left(\frac{Q_{rad,TE}}{Q_{rad,TM}}\right)^{-1}$.
Therefore, we can extract the photon generation ratio, $\Gamma_{TE}/\Gamma_{TM}$, from our experimental observables, namely $Q_{rad}$, $Q_{mat}$, and the accumulated CL signal, $N_{CL}$, in the relevant spectral peaks:
\begin{equation}
\frac{\Gamma_{TE}}{\Gamma_{TM}}=\frac{N_{CL, TE}}{N_{CL, TM}} \frac{Q_{rad}^{TE}+Q_{mat}}{Q_{rad}^{TM}+Q_{mat}}
\label{eq:ratio_Gamma_from_N_CL_and_Q}
\end{equation}
For the spectral peaks centered about 503 $nm$ and 518 $nm$ for an impact parameter $R_{sphere}$ (see Fig. \ref{fig:AR-spectrum}(e)), we find that $\Gamma_{TE}/\Gamma_{TM}=1.8 \pm0.1$. In the semi-analytical calculation of the WGM package the ratio of $\Gamma_{TE}/\Gamma_{TM}$ exhibits a sharp discontinuity at the sphere's boundary, dropping from $9.7$ to $ 1.1$ inside and outside of the sphere, respectively. Radiation into the TE-polarized mode dominates near the boundary since it is better confined \cite{BALAC2019,HeebnerJohnE2008}. 
It seems that our measured photon generation ratios include an inherent contradiction. On one hand, the electron energy is low, allowing only CL mediated by excitations \textit{in the material}. On the other hand, the measured generation rate ratio $\frac{\Gamma_{TE}}{\Gamma_{TM}}= 1.8$ befits CL emission right \emph{outside the sphere}, i.e., in vacuum, where the calculated value is $\frac{\Gamma_{TE}}{\Gamma_{TM}}=1.2$. In other words, the CL is emitted from material positioned in the vacuum. In reality, the sphere is imperfect, and may exhibit some ovality or surface roughness. Thus, the smearing of the interface between material and vacuum smooths the sharpness of the calculated transition of $\frac{\Gamma_{TE}}{\Gamma_{TM}}$. However, this seeming contradiction verifies that the CL generation is located right at the sphere's boundary. This surface sensitivity is granted here due to the edge-localization of the optical WGM, their edge field-discontinuities, and the spectral separability of the TE- vs. TM-polarized modes. Typically, CL is dominated by the bulk, where the optical density of states at the surface is small, making surface contribution negligible \cite{Mignuzzi2018,Coenen2017,sapienza2012}. The surface sensitivity is further enhanced by the geometry of the near-edge electron incidence. When the electron trajectory is tangent to the sphere ($r_e \lesssim R_{sphere}$) it is parallel to the surface and remains superficial over hundreds of nanometers. Thus, the surface propensity of the electron incidence and the photonic modes enable the selective sensing of surface phenomena via CL, without resorting to low electron energies \cite{kanaya1972}. 
This highlights the role of geometry and fabrication precision in shaping the CL response and sets the stage for engineering mode selectivity, polarization control, and directional output through careful tuning of the beam parameters and resonator dimensions.

\section{Conclusions}
In conclusion, we investigated cathodoluminescence in silica microspheres acting simultaneously as resonant cavities for optical whispering-gallery modes and as collimating lenses for external emission. While the various contributions to the CL are inherently convolved, we disentangle them by analyzing their spectral, spatial and angular resolved measurements. The properties of on-sample lensing of the sphere for CL originating in its vicinity are characterized according to its angular directionality. For the non-directional CL from WGM, we apply a spectral analysis to disentangle photon generation, radiative out-coupling, and absorptive loss mechanisms. Specifically, we find the quality factor related to material absorption and radiative whispering-gallery loss, $Q_{mat}$ and $Q_{rad}$, respectively. This decomposition is used to quantify photon generation rates and deduce the surface sensitivity of WGM CL. 
Altogether, our results establish silica microspheres as model systems to probe electron-photon interactions with mode selectivity, while also serving as practical elements for improving CL signal in advanced spectroscopy, exemplified by two particularly helpful aspects of nanoscopic photonics. First, it offers a path to investigate CL of surface electronic excitations by combining the edge-parallel electron irradiation and surface photonic modes. Second, the on-sample and high-NA collimation by the sphere shapes CL into a low-divergence and nearly diffraction-limited beam, and does so with high efficiency. Thus, it could enable low-NA and otherwise constrained CL-collection optics to produce a good spatial mode compatible with fiber coupling or quantum measurements\cite{chonan2014,kfir2021,tziperman2025two}. 

\section{Acknowledgment}
We gratefully acknowledge Hugo Lourenço Martins (CEMES-CNRS), Mohammad Joubat (TAU) for their help, and the Center for Light–Matter Interaction at Tel Aviv University for their support.
This research was supported by The Israel Science Foundation (grant No. 2992/24 and 1021/22).
O.K. gratefully acknowledges funding from the Gordon Foundation, and from the Young Faculty Award from the National Quantum Science and Technology program of the Israeli Planning and Budgeting Committee.
H.A. gratefully acknowledges funding from the National Quantum Science and Technology Program, the Center for Quantum Science and Technology at Tel-Aviv University and the French NanoX project (ANR-17-EURE-0009).

\bibliographystyle{apsrev4-2}
\bibliography{References_clean}

\end{document}

\maketitle

\begin{figure}[!htbp]
\centering
\noindent
\includegraphics [width=\columnwidth] {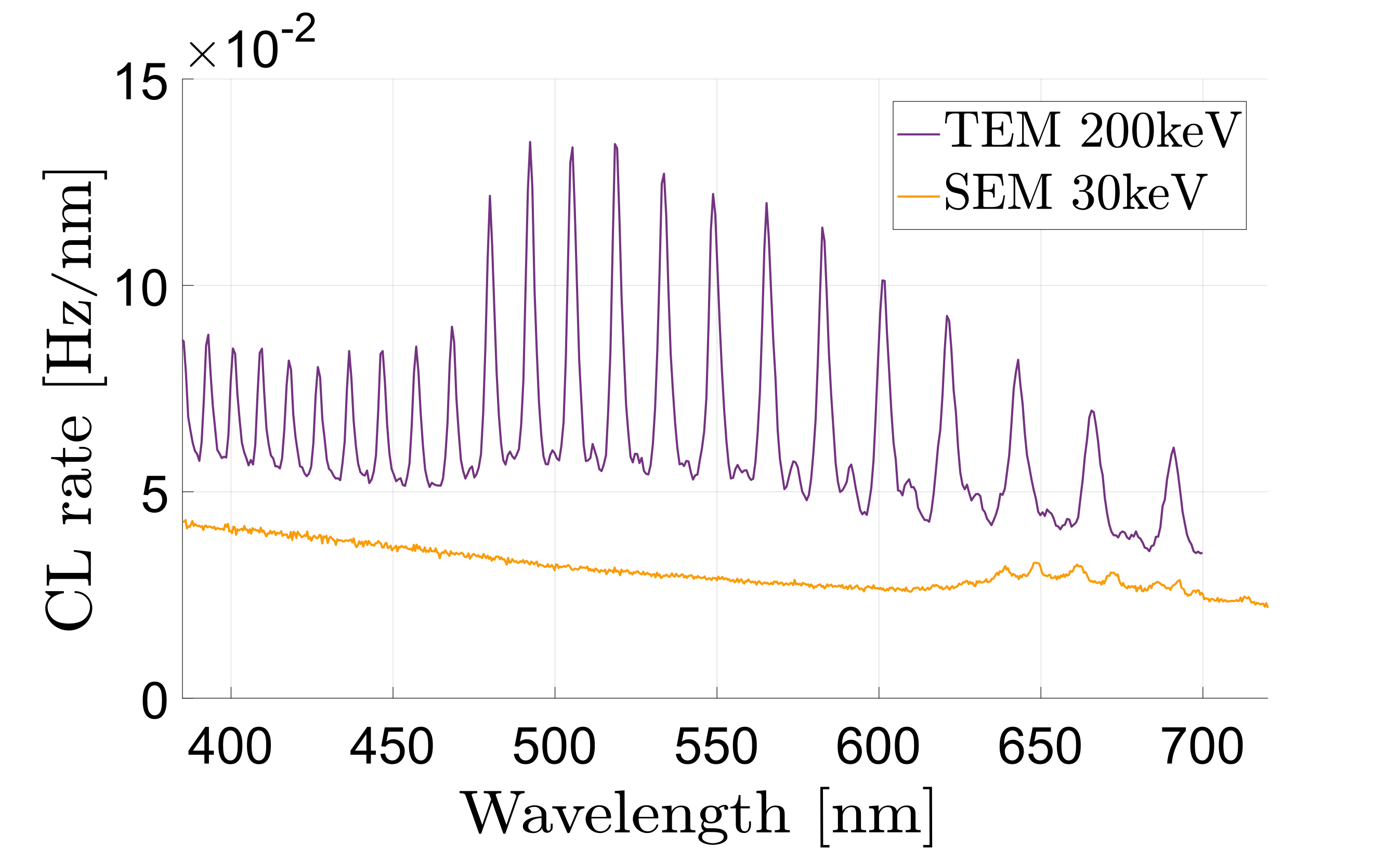}
\caption{CL spectra count rate extracted from 4.4 $\mu m$ silica glass sphere under 200keV electron beam excited in TEM (purple) and 30 keV in SEM (orange).} 
\label{fig:4umCL_30_200}
\end{figure}

\noindent
Figure \ref{fig:4umCL_30_200} presents the cathodoluminescence (CL) spectrum of a 4.4 $\mu m$ diameter silica glass sphere, acquired with a 30-second integration time under two conditions: 200 keV in the HF2000 TEM (Hitachi, purple) and 30 keV in the Apreo 2S-LV SEM (Thermo Fisher Inc., orange).

\begin{figure}[!htbp]
\centering
\noindent
\includegraphics [width=\columnwidth] {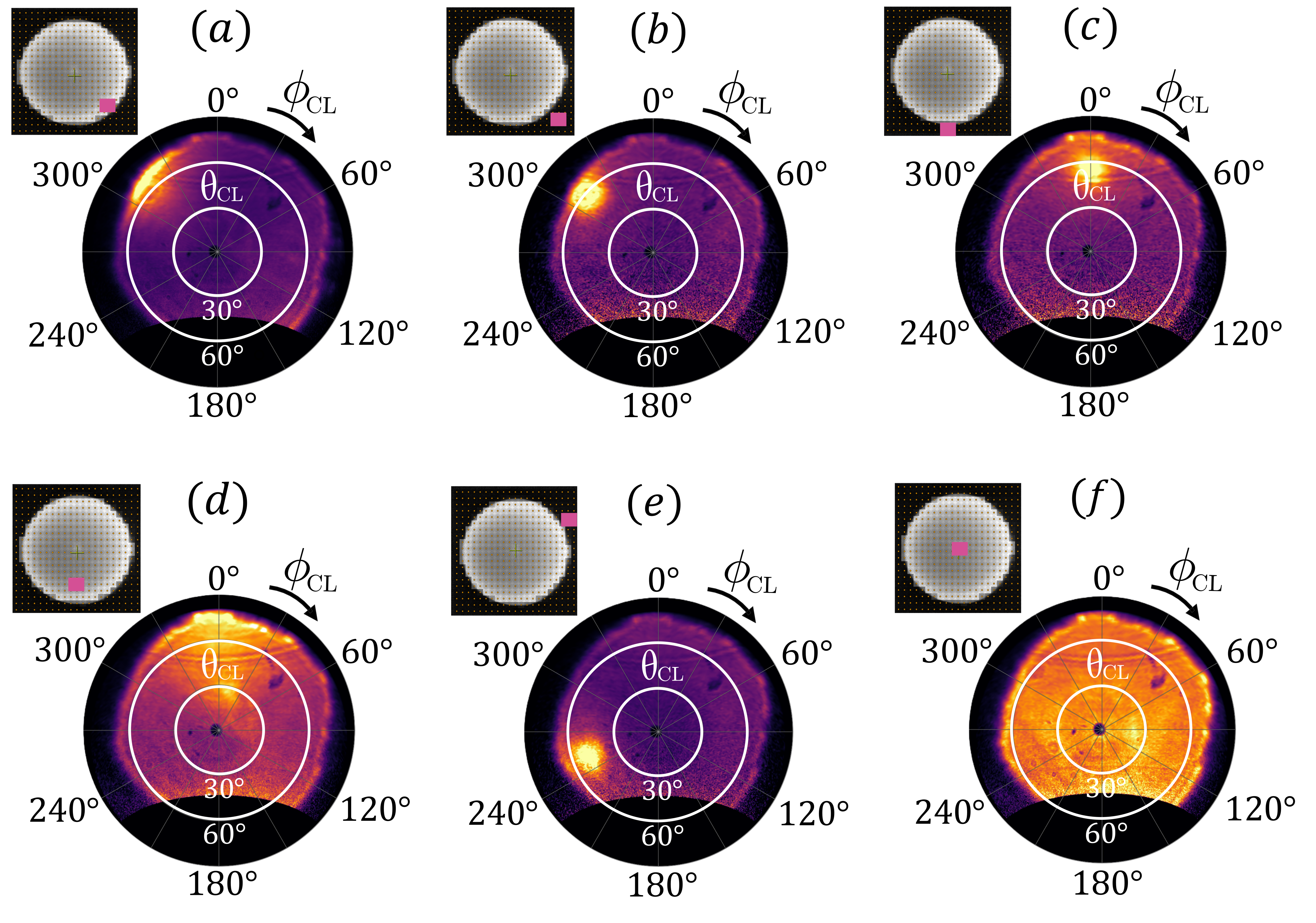}
\caption{(a)-(f) 2D maps showing the CL angular distribution from 2 $\mu m$ diameter silica sphere at distinct electron excitation position, marked with pink square in each inset.} 
\label{fig:AR_data}
\end{figure}

\newpage
\noindent 
Figure \ref{fig:AR_data} presents a matrix measurement over $21 \times 20$ positions, where at each location the angularly resolved CL was recorded with a 10-second integration time and a 50 $nm$ spectral filter centered at 550 $nm$. Fig.\ref{fig:AR_data}(a)-(f) show six representative examples from the matrix measurement.

\begin{figure}[!htbp]
\centering
\noindent
\includegraphics [width=\columnwidth] {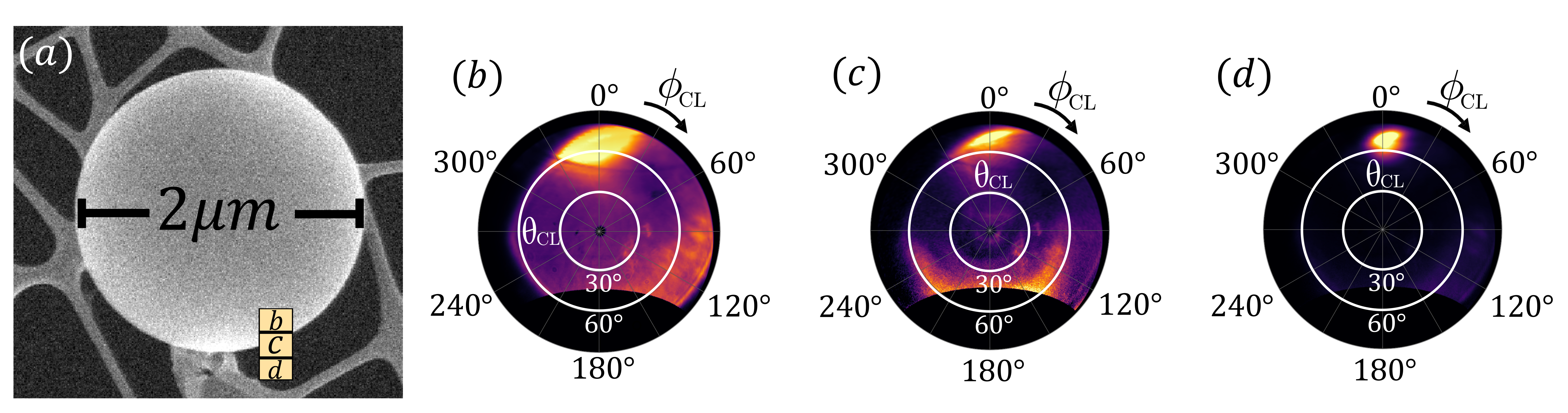}
\caption{(a) CL angular distribution measured from a 2 $\mu m$ diameter silica sphere at three excitation positions aligned in a row, indicated by yellow squares in (a) and labeled (b–d). (b)-(d) show the corresponding 2D angular maps for each excitation position.} 
\label{fig:AR_row_data}
\end{figure}

\newpage
\noindent
Figure \ref{fig:AR_row_data} displays the angular distributions obtained at three consecutive excitation positions in a row, recorded without a spectral filter and with a 120-second integration time.

\begin{figure}[!htbp]
\centering
\noindent
\includegraphics [width=\columnwidth] {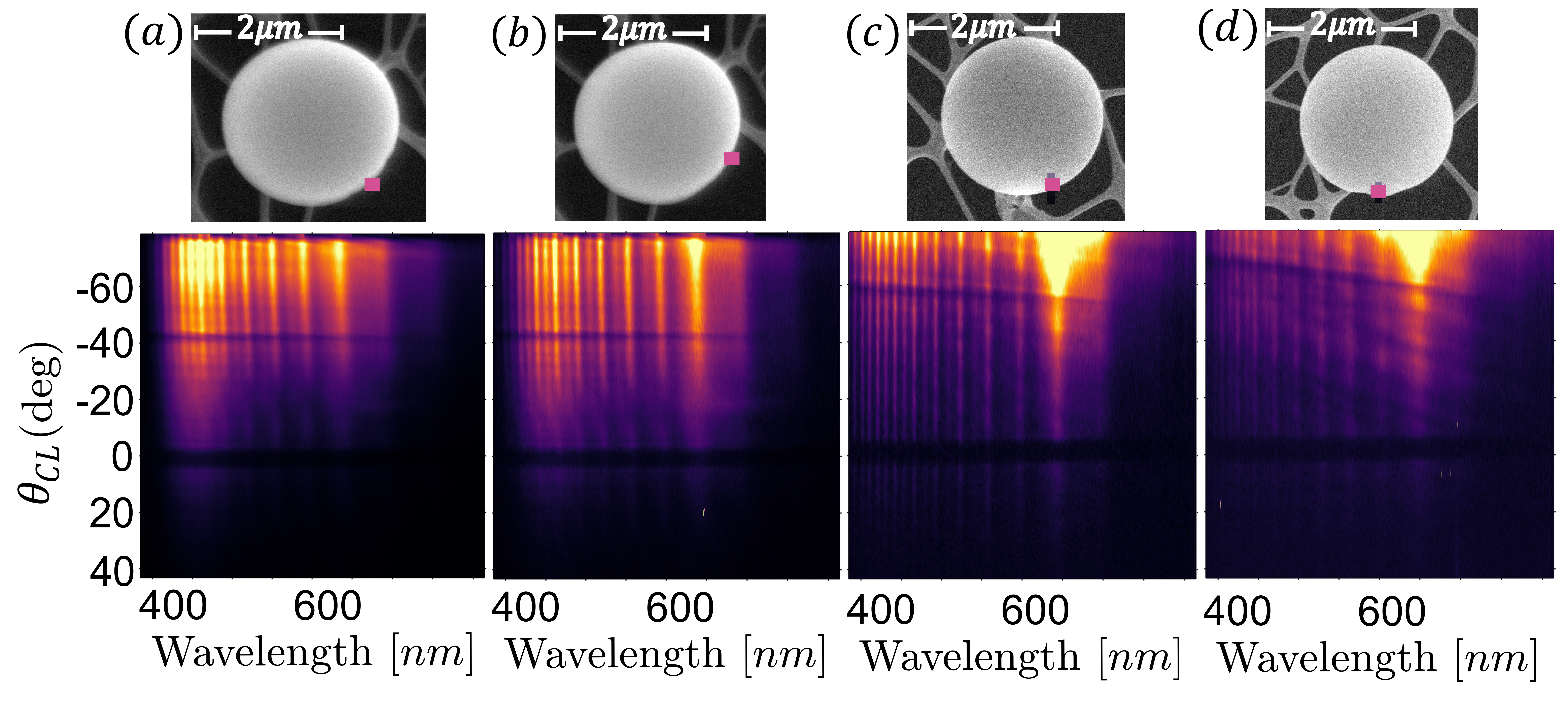}
\caption{$\theta$-resolved CL spectrograms from 2 $\mu m$ diameter silica spheres for different electron excitation positions, indicated by pink squares in the insets. (a–b) spectrograms from the same 2 $\mu m$ sphere at two distinct excitation positions. (c–d) spectrograms from two different 2 $\mu m$ spheres, each with a distinct excitation position. All plots exhibit WGM peaks with spatially varying backgrounds.} 
\label{fig:ARSpect_data}
\end{figure}

\hfill \break
\noindent
Figure \ref{fig:ARSpect_data} presents $\theta$-resolved CL spectrograms of 2 $\mu m$ diameter silica spheres, acquired with a 60-second integration time at different electron excitation positions.

